\documentclass{iau} 
\usepackage{graphicx}
\usepackage{rotating}

\title[Dichotomy of radio loud and radio quiet quasars in four dimensional eigenvector one (4DE1) parameter space] 
{Dichotomy of radio loud and radio quiet quasars in four dimensional eigenvector one (4DE1) parameter space \footnote{Based on observations obtained at the CAHA Observatory, Calar Alto, Spain}} 

\author[Shimeles Terefe \& Ascensi\'{o}n Del Olmo Orozco \& Paola Marziani]   
{Shimeles Terefe$^1$,
 Ascensi\'{o}n Del Olmo Orozco$^2$, Paola Marziani$^3$ \and Mirjana Povi\'{c}$^{1,2}$}
\affiliation{$^{1}$Ethiopian Space Science and Technology Institute (ESSTI), Addis Ababa, Ethiopia\\ email: shimeles11@gmail.com  \\[\affilskip]
$^2$Instituto de Astrofisica de Andaluc\'{i}a (IAA-CSIC), Granada,
Spain  \\
[\affilskip]
$^3$Istituto Nazionale di Astrofisica (INAF), Osservatorio Astronomico 
di Padova, Padova, Italy \\
[\affilskip]
}
\pubyear{2019}
\volume{356}  
\jname{Nuclear Activity in Galaxies Across Cosmic
Time}
\editors{M. Povi\'c, P. Marziani, J. Masegosa, H. Netzer,\\ S. H. Negu, \&
	S. B. Tessema, eds.}
\begin{document}

\maketitle

\begin{abstract}
Recent work has shown that it is possible to systematize quasars (QSOs) spectral diversity in 4DE1 parameter space. The spectra contained in most of the surveys have
low signal to noise ratio which fed the impression that all QSO’s are spectroscopically similar. Exploration of 4DE1 parameter space gave rise to the concept of two populations of QSOs that present important spectroscopic differences. We aim to quantify broad emission line differences between radio quiet and radio loud sources by exploiting more complete samples of QSO with spectral coverage in H${\beta}$, MgII and CIV emission lines. We used a high redshift sample (0.35 $<$ z $ <$ 1) of strong radio emitter QSOs observations from Calar Alto Observatory in Spain. 
\end{abstract}

\firstsection 
\section{Introduction}

Over fifty years after their discovery, people are beginning to see progress in both
defining and contextualizing the properties of QSOs, some of the brightest AGNs
(\cite[Sulentic et al. 2007]{Sulentic et al. 2007}). Bright type-1 AGNs show widely differing line profiles, intensity ratios and ionization levels (Marziani et al. 2018). 

A much debated problem in AGN studies involves the
possibility of a real physical dichotomy between radio loud (RL) and radio quiet (RQ) QSOs (Zamfir et al. 2008).
Another complication is introduced by the fact that some good fraction of RQ sources
share common properties with the RL QSOs as discussed in Zamfir et al. (2008). For
instance, about 30\,-\,40\,\% of RQ QSOs are spectroscopically similar to RL (Sulentic et al. 2000) and with the improvement of radio interferometry techniques, it was possible to notice that both QSO types are capable of producing radio jets (Chiaberge, $\&$ Marconi 2011).

The other which remains a perplexing question 50 years after the discovery of QSO is the origin of radio loudness (Ruff, 2012). From the theoretical point of view, in spite of the great advancement in the ability of
collecting unbiased sets of data, most of the researchers argued, the origin of the relativistic radio jets in AGNs as an open question.

A first contextualization, RQ vs. RL (Sulentic et al. 1995, Corbin 1997) showed intriguing spectroscopic differences between the two types of QSOs with large blue shifts observed in the emission line profiles only among RQ sources.

Recent work has shown that it is possible to systematize quasar spectral diversity in a
space called 4DE1 parameter space (see Marziani et al. 2018 and references therein). As
stated in Zamfir et al. (2008), the value of studying the RL phenomenon within the 4DE1
context is at least two fold: (i) the approach compares RL and RQ sources in a parameter space defined by measures with no obvious dependence on the radio properties (Marziani et
al. 2003b), (ii) it allows to make predictions about the probability of radio loudness for
any population of QSOs with specific optical (UV) spectroscopic properties. As the RQ/RL separation in 4DE1 is
not complete (Zamfir et al. 2008), many open questions are present after 50+ years of
study of QSOs.

Therefore, this work studies the properties of RL and RQ QSOs by using the 4DE1 parameter space. We specifically consider a possible dichotomy between them, and the reason behind observed low fractions of RL.

\section{Data}
The work focus on type 1 QSOs,
as they can be unambiguously identified based on the presence of a broad component in
the hydrogen Balmer emission lines (mainly in H$\beta$$\lambda4861$), the doublet of MgII$\lambda$2800, or
in High Ionization Lines (HILs) like CIV$\lambda$1549 in the UV. The best clues to study the RQ/RL
dichotomy in QSOs lies in the optical/UV spectra and many of the brightest RL QSOs with radio
coverage have no published optical spectra with S/N high enough to permit a detailed study. Data from astronomical archives were supplemented with new data obtained at the Observatory of Calar Alto in Spain. Regarding the new data, a sample of 50 strongly-RL QSOs were obtained by using 
the TWIN spectrograph attached at the 3.5m telescope of the Calar Alto Observatory. The TWIN spectrograph has two arms that allows to obtain (1) near UV spectra for studying the region of the MgII$\lambda$2800 doublet, and (2) the H$\beta$-FeII
region for the QSOs in the redshift range (0.35 $<$ z $<$ 1).

Radio images and the fluxes at radio
continuum were obtained from the
VLA Faint Images of the Radio Sky
at Twenty-Centimeters (FIRST) survey
and from the NVSS survey for those
QSOs with no available FIRST radio
measurements. This information is fundamental to identify QSOs according to their
morphology at radio frequencies and to determine the
power of radio emission.

The Kellermann factor $R_{k}$ which is a value defined by the ratio between radio/optical flux density was used as a discriminant between RQ and RL. High S/N spectra with H$\beta$, CIV$\lambda$1549 and MgII$\lambda$2800 coverage were considered.

\section{Data analysis and preliminary results}
Analyzing the broad lines by doing multicomponent non-linear
fitting of broad emission lines in particular of H$\beta$, FeII, MgII and the
UV lines of CIV and HeII allows us to quantify the properties and
kinematics of the broad line region, to detect very broad
components (observed only in population B of QSOs within the
4DE1 scheme) as well as to calculate fundamental quantities such
as the mass of the super massive black hole (SMBH) and the
Eddington ratio $(L/L_{Edd})$. In order to get spectra suitable for scientific use, we used a standard spectroscopic data reduction by using IRAF astronomical package which produces a spectrum as shown in Figure 1 left for one of our source, S5$\_$1856+73 at z\,=\,0.46 which incorporates most of the emission lines.  

The optical (H$\beta$-FeII) and UV (MgII) analysis mainly focus on the spectral fitting of emission lines using the SPECFIT routine in the IRAF package. The result of the SPECFIT for H$\beta$ emission line for S5$\_$1856+73 is shown in Figure 1 right.

In radio, we searched the VLBA radio map  and the NVSS contour map as shown in Figure 2 for S5$\_$1856+73  which can be taken as an indication of very different morphologies in radio. The analysis of the connection between the optical/UV and radio measurements will be presented in a forthcoming paper. 
\begin {figure}[h!] 
\centering
\includegraphics [width=2.5in,height=1.70in ]{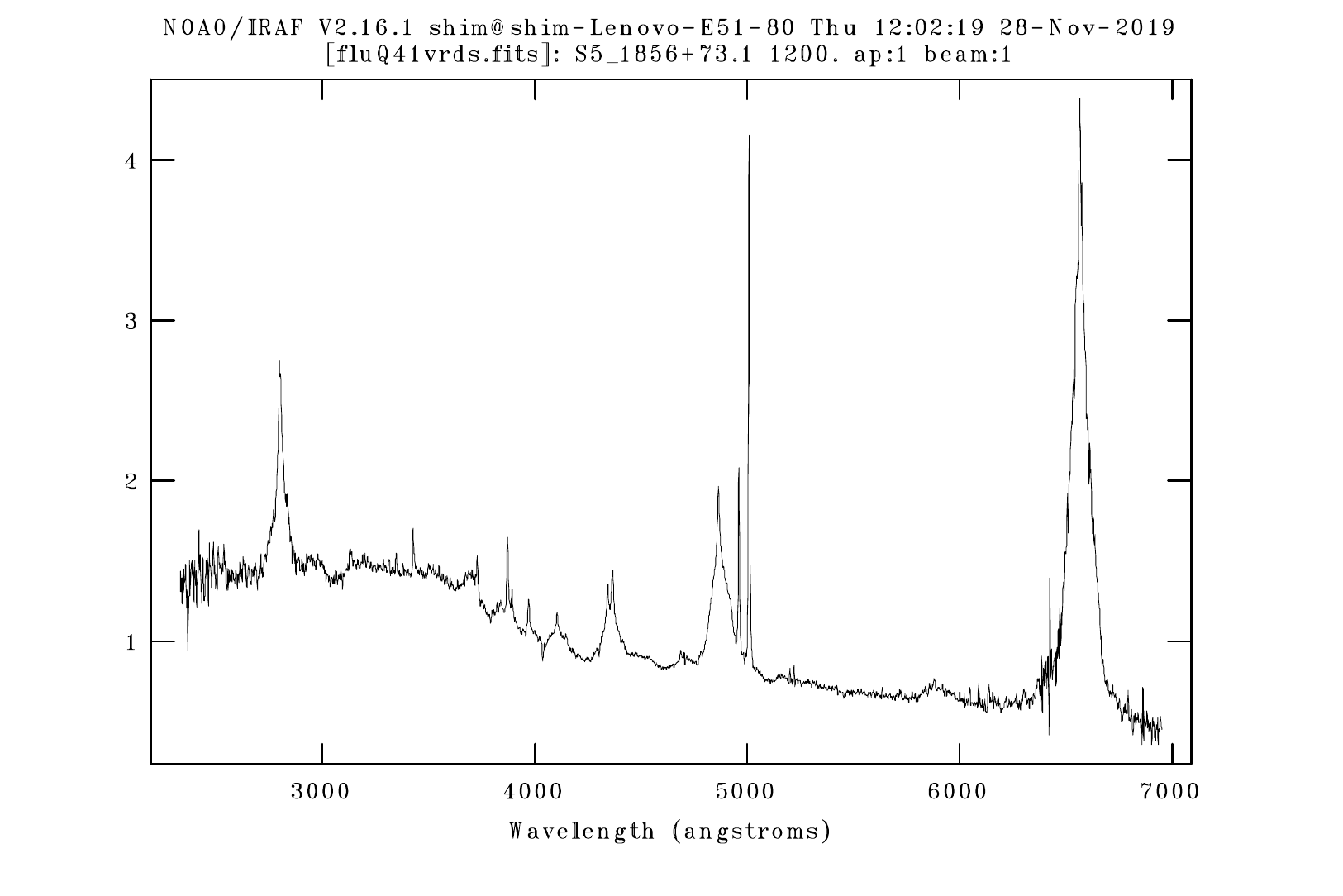}~~~~~~~~
\includegraphics [width=2.25in,height=1.60in ]{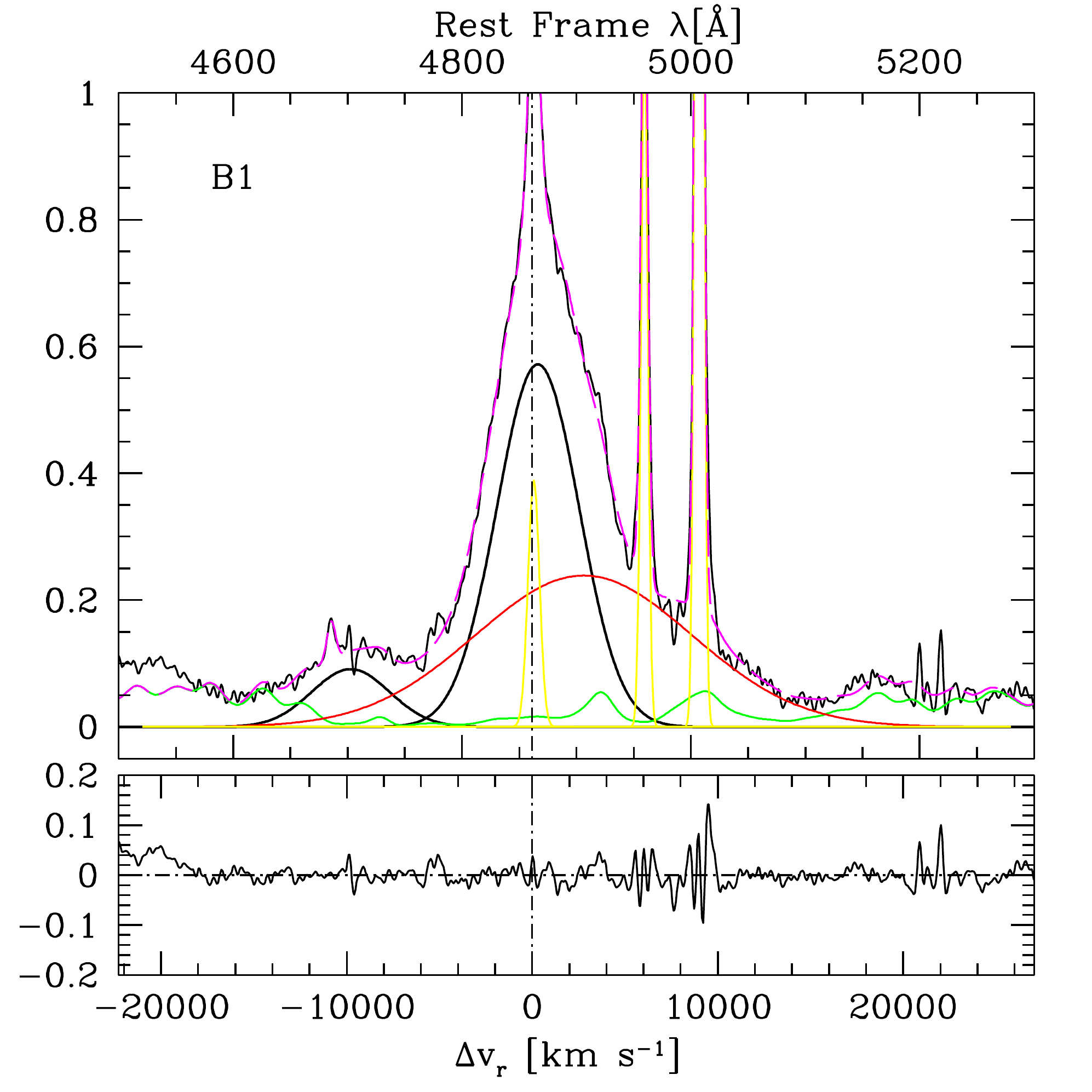}
\caption{ (Left) Rest frame spectra for S5$\_$1856+73, abscissa corresponds to rest frame wavelength, ordinate corresponds to specific flux
in units of $10^{-15}$erg $s^{-1} cm^{-2} \AA^{-1}$ and (right) multicomponent fitting in
the H$\beta$ region. Yellow lines represent
the narrow components of [OIII] and H$\beta$; black line for the broad component; red
line corresponds to the very broad
component and the green one the FeII.
Bottom plot corresponds to
the residuals of the fitting.} 
\label {fig:p}
\end {figure}
\begin {figure}[h!]
\centering
\includegraphics [width=2.in,height=1.55in ]{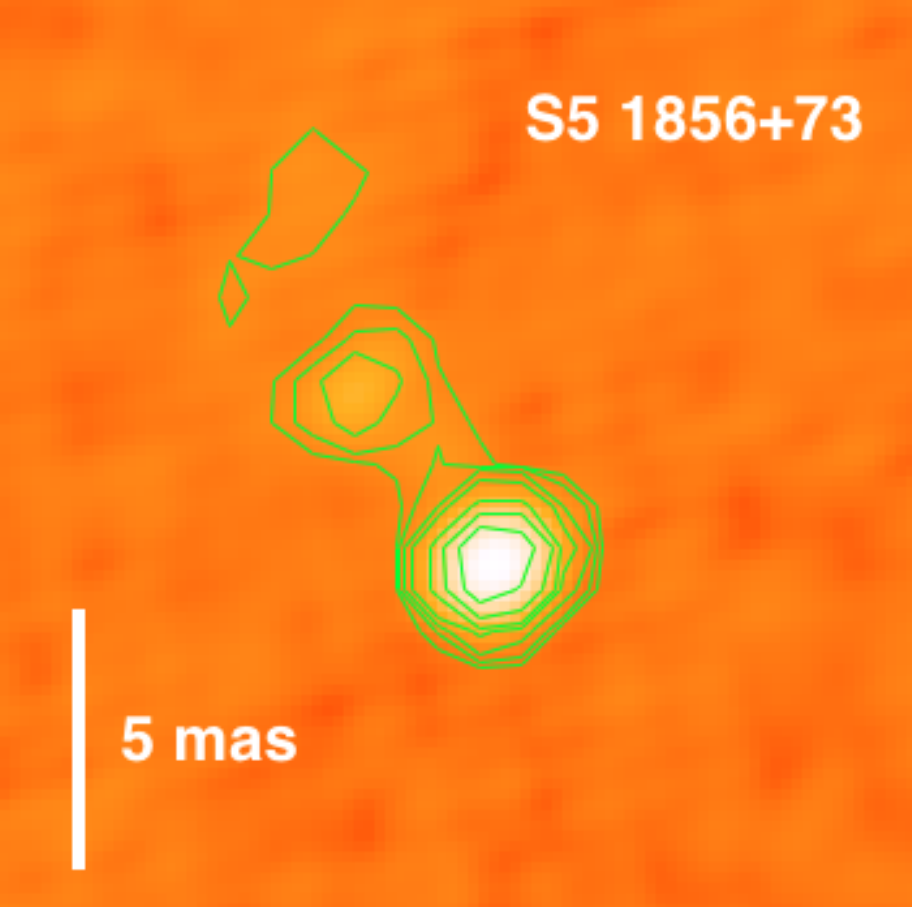}~~~~~~~~~~~~
\includegraphics [width=2.in,height=1.60in]{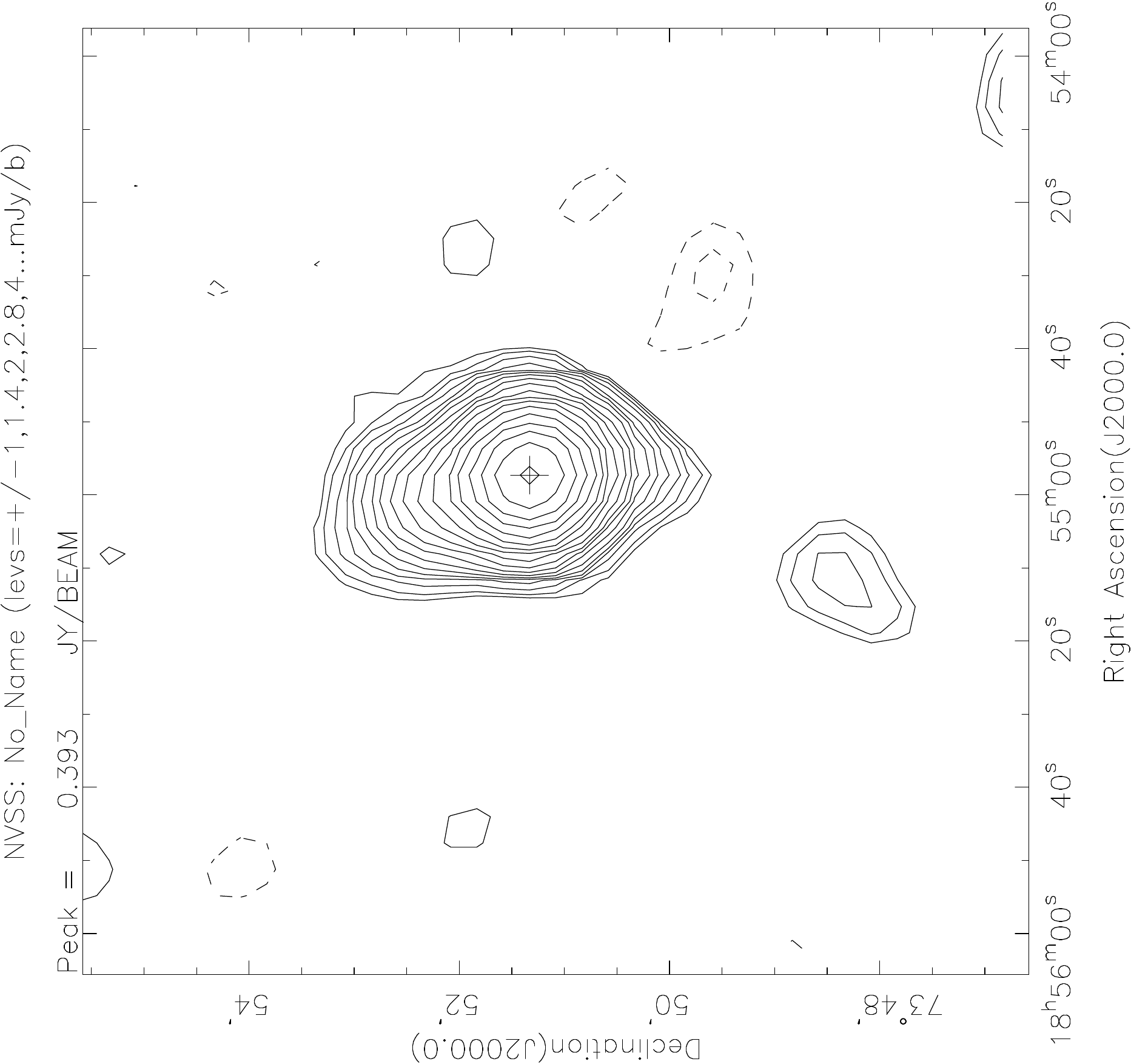}
\caption{  (left) VLBA radio map of S5$\_$1856+73  and its NVSS contour map (right)} 
\label {fig:p}
\end {figure}
\section*{Acknowledgment}
AdO acknowledges financial support from Spanish grants AYA2016-76682-C3-1-P and SEV-2017-0709. We also acknowledge ESSTI and Jimma University for their support.

\end{document}